\documentclass[11pt]{article}%{llncs}

\usepackage[latin1]{inputenc}
\usepackage[american]{babel}

\usepackage{fp, calc}
\usepackage[margin = 3.20cm]{geometry}

\usepackage{amsmath}
\usepackage{amssymb}
\usepackage{xspace}
\usepackage{theorem}
\usepackage{graphicx}
\usepackage{subfig}
\usepackage{epsfig}
\usepackage{ifpdf}
\usepackage{url,hyperref}
\usepackage{latexsym}
\usepackage{euscript}
\usepackage{xspace}
\usepackage{color}
\usepackage{makeidx}
\usepackage{picins,wrapfig}
\usepackage{stackrel}
\usepackage{amscd}
\usepackage[all]{xy}
\usepackage{multirow}
\usepackage{lineno}
\usepackage{algorithmic}
\usepackage{algorithm}

\usepackage{latexsym}
\usepackage{amsmath}
\usepackage{amssymb}
\usepackage{color}
\usepackage{graphicx}
\usepackage{algorithmic}
\usepackage{algorithm}

\newcommand{\norm}[1]{\| #1 \|}

\newcommand{\R}{\mathbb R}

\newcommand{\rch}{{\rm rch}}

\newcommand{\reel}{{\mathbb R}}

\newcommand{\dist}[2]{d(#1,#2)}

\newcommand{\man}{\mathcal{M}}
\newcommand{\M}{\mathcal{M}}
\newcommand{\lfs}{\mathrm{lfs}}

\newenvironment{proof}[1][{}]{%\novbskip%
  \begin{trivlist}\item[]\textit{Proof #1}\quad}%
  {\hfill\hspace*{\fill}~$\square$\end{trivlist}}

\newtheorem{theorem}{Theorem}%[section]

\newtheorem{lemma}[theorem]{Lemma}

\usepackage{url,hyperref}

\usepackage{wrapfig}

\title{
An elementary approach to tangent space variation\\
on Riemannian submanifolds
\thanks{
A weaker version of this
result first appeared in Chapter~4 
of the third author's PhD thesis~\cite[Lem.~4.3.2]{ghosh:hal-01095861}.
}
}

\date{}

\author{
Jean-Daniel Boissonnat
\footnote{
Geometrica, INRIA, Sophia Antipolis, France
\url{Jean-Daniel.Boissonnat@inria.fr}
}
\and
Ramsay Dyer
\footnote{
Johann Bernoulli Institute,
University of Groningen, 
Groningen, The Netherlands,
\url{r.h.dyer@rug.nl}
}
\and
Arijit Ghosh
\footnote{
D1: Algorithms \& Complexity,
Max-Planck-Institut f\"ur Informatik,
Germany
\url{agosh@mpi-inf.mpg.de}
}
}

\begin{document}

\maketitle

\begin{abstract}

We give asymptotically tight estimates of tangent 
space variation on Riemannian submanifolds of Euclidean space
with respect to the local feature size of the submanifolds.
We show that the result follows directly from structural 
properties of local feature size
of the Riemannian submanifold and some elementary Euclidean geometry.
We also show that using the tangent variation result one can 
prove a new structural property of local feature size function. 
This structural property 
is a generalization of a result of Giesen and 
Wagner~\cite[Lem.~7]{DBLP:journals/dcg/GiesenW04}.

\paragraph{Keywords.} Local feature size, Lipschitz function,
Riemannian manifold and tangent space.
\end{abstract}

\section{Introduction}

Let $\man$ be a compact Riemannian submanifold of the Euclidean 
space $\reel^{N}$. For a point $p$ in $\man$, we  denote by 
$T_{p}\man$ the tangent space of $\man$ at $p$. 

In this note 
we prove the following result:
\begin{theorem}[Tangent space variation]\label{theorem-tangent-variation-submanifolds}
Let $p,\, q \in \man$ and $\|p-q\| = t \lfs(p)$.
\begin{description}
\item[(i)]
If $t \leq 1/4$ then
\begin{equation}
    \sin \angle (T_{p}\man, \, T_{q}\man)  \leq  
%\min \{ 1, \, t \,f(t)\}, 
t f(t), \quad
      \mbox{where }  f(t) = \frac{(2+3t+2t^{2})^{2} + 4t + 5}{2-2t}.
%    = \frac{9}{2} + O(t).
\end{equation}
Observe that $f(x) = \frac{9}{2} + O(t)$, and
if $t \leq \frac{1}{10}$ then $f(t) < 6$.
We will define $\lfs(\cdot)$ in Section~\ref{sec:medial-axis-lfs}.

\item[(ii)] {\bf Better asymptotic constant:}
 If $t \leq \frac{19}{200}$, then
 \begin{equation}
    \sin \angle (T_{p}\man, \, T_{q}\man)  \leq  
%\min \{ 1, \, t \,g(t)\}, 
%tg(t), \quad \mbox{where }   g(t) = 3 + O(t).
3t + O(t^2).
 \end{equation}
\end{description}
\end{theorem}
%We will define $\lfs()$ in the next section. 

%\begin{remark}[Imroved asymptotic bound]
% In Theorem~\ref{theorem-tangent-variation-submanifolds-new} 
% we will show that with a little bit more 
% effort we can improve the asymptotic constant in 
% Theorem~\ref{theorem-tangent-variation-submanifolds} 
% from $\frac{9}{2}$ to $3$. 
%\end{remark}

A special case of this result in the context of $2$-dimensional
Riemannian submanifolds in $\reel^{3}$ was proved by Amenta and
Dey~\cite[Thm.~2]{DBLP:journals/corr/AmentaD14}, and a weaker bound
for the general case of Riemannian submanifolds of Euclidean space was
proved by Niyogi, Smale and Weinberger~\cite[Prop.~6.2 \&
6.3]{nsw-fhswhc-04}, using a global bound on the local feature size
function, rather than its local value.
Belkin, Sun and Wang~\cite[Lem.~3.4]{DBLP:conf/soda/BelkinSW09} 
showed that using cosine-law in the proof of the result of 
Niyogi, Smale and Weinberger one can show an optimal $O(t)$ variation
bound.

Our result is more general than these previous results, and we show
that it is asymptotically tight. Our proof is elementary and follows
directly from structural properties of the {\em local feature size function},
$\lfs(\cdot)$, of Riemannian submanifolds and some elementary Euclidean geometry.

We expect that the result of Niyogi et al.~\cite[Prop.~6.2 \&
6.3]{nsw-fhswhc-04} together with 
Belkin et al.'s improvement~\cite[Lem.~3.4]{DBLP:conf/soda/BelkinSW09} 
can be extended to the case of local feature size. 
Our main contribution is the simplicity of the proof of the 
tangent variation result wrt to the local feature size. 
We also show that using the tangent variation result one can 
prove a new structural property of local feature size, 
see~Lemma~\ref{lem:improved-sampling-2}. This result 
is a generalization of Lemma~7 from~\cite{DBLP:journals/dcg/GiesenW04}.

\paragraph{Notation.}
We denote the standard Euclidean distance between
$x , \, y \in \reel^{N}$ by $\|x-y\|$, and for a point
$x \in \reel^{N}$ and a set $X \subseteq \reel^{N}$, the distance
between $x$ and $X$ will be denoted by
$\dist{x}{X} = \inf_{y \in X} \|x-y\|$.

For $x \in \reel^{N}$ and $r \geq 0$, we denote balls and spheres by
$B(x,r) = \left\{ y \in \reel^{N} \, :\; \|x-y\| < r \right\}$,
$\overline{B}(x,r) = \left\{ y \in \reel^{N} \, :\; \|x-y\| \leq r
\right\}$, and $\partial B(p,r) = \overline{B}(p,r) \setminus B(p,r)$.

For $S \subseteq \reel^{N}$, we  use $B_{\mid S}(p,r)$ 
to denote $B(p,r)\cap S$. We similarly define 
$\overline{B}_{\mid S}(p,r)$ and $\partial B_{\mid S}(p,r)$.

If $U$ and $V$ are vector subspaces of $\R^{d}$, with
$\dim(U) \leq \dim(V)$, the {\em angle} between them is defined by
\begin{equation*}
  \angle (U,V) = \max_{\substack{u \in U, \\ \|u\|=1}} \;
  \min_{\substack{v \in V, \\ \|v\|=1}} \angle(u,v)
  = 
  \arccos \inf_{\substack{u \in U, \\ \|u\|=1}}\;
  \sup_{\substack{v \in V, \\ \|v\|=1}}
  \langle u , v \rangle .
\end{equation*}
This is the largest principal angle between $U$ and $V$.  It is easy
to show from the above definition that if $\dim(U) = \dim(V)$ then
$\angle (U, V) = \angle (V, U)$
(see~\cite[Lem.~2.1]{DBLP:journals/dcg/BoissonnatG14}).

The angle between affine subspaces is defined
as the angle between the corresponding parallel vector subspaces.

\section{Medial axis and local feature size}
\label{sec:medial-axis-lfs}

The {\em medial axis} of $\M$ is the closure of the set of points of
$\R ^N$ that have more than one nearest neighbor on $\M$.  The {\it
  local feature size} of $x\in \M$, $\lfs(x)$, is the distance of $x$
to the medial axis of $\M$~\cite{AmentaBern}. As is well known and can
be easily proved, $\lfs$ is {\it Lipschitz continuous}, i.e.,
$$
  \lfs(x) \leq \lfs(y)+\| x-y\|.
$$

Amenta and Dey~\cite[Thm.~2]{DBLP:journals/corr/AmentaD14}, 
proved the following tangent variation result 
for the case of $2$-dimensional Riemannian submanifolds of $\reel^{3}$:
\begin{theorem}[Two-dimensional case]
 Let $\man$ be $2$-dimensional Riemannian submanifold of $\reel^{3}$. 
 Let $p, \, q \in \man$ such that $\|p-q\| = t \, \lfs(p)$ 
 with $t \leq \frac{1}{3}$. Then 
 $\sin \angle (T_{p}\man, \, T_{q}\man) \leq \frac{t}{1-t}$.
\end{theorem}
The proof of the above result is restricted to the 
case where the dimension of the submanifold is two and the 
codimension is one.

The infimum of
$\lfs$ over $\M$ is called the {\em reach } $\rch(\man)$ of $\man$.  
%Federer~\cite{} proved that XXXXXXXX
%   
Niyogi, Smale and Weinberger~\cite[Prop.~6.2 \& 6.3]{nsw-fhswhc-04}
first proved the following bound on the tangent variation on Riemannian
submanifolds:
\begin{theorem}
  \label{thm:niyogi}
 Let $\man$ be a Riemannian submanifold of $\reel^{N}$. Let $p, \, q \in \man$
 such that $\| p - q |\ = t \, \rch(\man)$ with $t \leq 1/2$. Then
 $\sin \angle (T_{p}\man, \, T_{q}\man) \leq 2\sqrt{t(1-t)}$.
\end{theorem}
Belkin-Sun-Wang~\cite[Lem.~3.4]{DBLP:conf/soda/BelkinSW09} 
showed that using cosine-law in the proof of 
Theorem~\ref{thm:niyogi} one can get the following improvement:
\begin{theorem}\label{thm-Belkin-Sun-Wang-improvement-09}
 Let $\man$ be a Riemannian submanifold of $\reel^{N}$. Let $p, \, q \in \man$
 such that $\| p - q |\ = t \, \rch(\man)$ with $t \leq 1/2$. Then
 $\sin \angle (T_{p}\man, \, T_{q}\man) \leq 2\, t \sqrt{1-t^{2}}
  = (2-O(t^{2}))t$.
\end{theorem}

The proof of Theorem~\ref{thm:niyogi} uses tools from differential geometry,
such as {\em parallel transport}. Note that $O(t)$ bound in 
Theorem~\ref{thm-Belkin-Sun-Wang-improvement-09} is tight,
see Section~\ref{sec:discussion}.
But, as we have already mentioned, the result 
use a global bound on the local feature size function, i.e. the reach,
rather than its local value. 
We expect that this result can
be extended to the case of local feature size 
function with more work connecting local feature size 
to other intrinsic properties of the submanifolds like
the {\em strong convexity radius}.
Our proof, on the other hand,
is both elementary and works directly for 
local feature size function with no assumption on the global 
bound on the local feature size function.

\section{Proof of Theorem~\ref{theorem-tangent-variation-submanifolds}~(i)}
\label{sec:the-proof}

The following lemma, proved in~\cite[Lem.~6 \&
7]{DBLP:journals/dcg/GiesenW04}, states some basic properties of local
feature size function.
\begin{lemma}\label{lem:sampling}
   \begin{enumerate}
    \item
	Let $p,\, q \in \man$ such that
	$\| p-q \| = t\,\lfs(p)$ with $t <1$, then
	$\dist{q}{T_{p}\man} \leq \frac{t^{2}}{2} \lfs(p)$.

    \item
	Let $p \in \man$ and $x \in T_{p}\man$ such that
	$\| p-x \| \leq t\,\lfs(p)$ with $t \leq \frac{1}{4}$,
	then 
	$\dist{x}{\man} \leq 2t^2\,\lfs(p)$.
   \end{enumerate}
\end{lemma}
We prove
Theorem~\ref{theorem-tangent-variation-submanifolds} using the 
above result.

%From Lemma~\ref{lem-affine-space}, we have $\angle (T_{p}\M, T_{q}\M)
%= \angle (T_{q}\M, T_{p}\M)$.

Let $t = \frac{\| p - q \|}{ \lfs(p)}$ with $t \leq 1/4$. 
Using the fact that $\lfs$ is $1$-Lipschitz, we have
\begin{equation}\label{eqn-lfs-p-q-t}
	(1-t) \, \lfs(p) \; \leq \; \lfs(q) \; \leq \; (1+t)\, \lfs(p)
\end{equation}

%For a unit vector $u$ in $T_{p}\M$, let $p_{u}$ be a point in $T_{p}\M$ such that

Let $u$ be an unit vector in $T_{q}\M$.  Let
$q_{u} = q + t\, \lfs(q)\cdot u$, and let $q'_{u}$ denote the point
closest to $q_{u}$ on $\M$.  Then, from Lemma~\ref{lem:sampling}~(2),
we have $\|q_{u}-q_{u}'\| \leq 2t^{2}\lfs(q)$.  Therefore, using the
fact that $\lfs(q) \leq (1+t)\,\lfs(p)$ (see
Eq.~\eqref{eqn-lfs-p-q-t}), we have
\begin{eqnarray*}
	\|p- q'_{u}\|  &\leq& \| p - q \| + \| q - q_{u} \| + \| q_{u} - q'_{u} \|\\
	 &\leq& t\, \lfs(p) + (t+2t^{2})\lfs(q) \\
	 &\leq& t\, \lfs(p) + (t+2t^{2})(1+t)\lfs(p) \\
	 &=& t \left( 2 + 3t + 2t^{2}\right) \lfs(p).
\end{eqnarray*}
	
%Using Lemma~\ref{lem:sampling}~(2) and Eq.~\eqref{eqn-lfs-p-q-t}, we have
%\begin{eqnarray}\label{eqn-dist-tq-p}
%	\dist(p, T_{q}\M) &\leq& \|p-q\|
%	\sin \angle (pq, T_{q}\M) \nonumber\\
%	&\leq& \frac{t^{2}}{2(1-t)^{2}} \lfs(q)
%\end{eqnarray}
Using Lemmas~\ref{lem:sampling}~(1) and (2), and Eq.~\eqref{eqn-lfs-p-q-t}, we have
\begin{eqnarray}\label{eqn-dist-tq-pu}
	\dist{q_{u}}{T_{p}\M}  
	&\leq&  \dist{q'_{u}}{T_{p}\M} + \norm{q_{u} - q'_{u}} \nonumber\\
	&\leq& \frac{t^{2}}{2}\left( 2+3t+2t^{2}\right)^{2} \lfs(p) + 2t^{2}\lfs(q) \nonumber\\
	&\leq& \frac{t^{2}}{2}\left( 2+3t+2t^{2}\right)^{2} \lfs(p) + 2t^{2}(1+t)\lfs(p) \nonumber\\
%	&\leq& \frac{t^{2}}{2}\left( \frac{2+t}{1-t} \right)^{2}\lfs(q) + 2t^{2}\frac{\lfs(q)}{1-t}
	&=& \frac{t^{2}}{2}\left( (2+3t+2t^{2})^{2} + 4(1+t) \right) \lfs(p)
\end{eqnarray}
%Let $\eta = \max \{ \dist(p,T_{q}\M),\, \dist(p_{u},T_{q}\M)\}$. From Eq.~\eqref {eqn-dist-tq-p} and
%\eqref{eqn-dist-tq-pu}, we have
%$$\eta \leq \frac{2t^{2}}{1-t} \left( \frac{(1+t)^{2}}{(1-t)} + 1\right) \lfs(q)$$
	
%	Since $e_{u}$ is an edge, hence $\Theta_{e_{u}} = 1$.

%From Lemma~\ref{lem:sampling}~(1), 
%there exists  unit vector $v$ in $T_{q}\M$ such that

\begin{figure}%[h]
%  \vspace{-25pt}
  \begin{center}
    \includegraphics[width=0.50\textwidth]{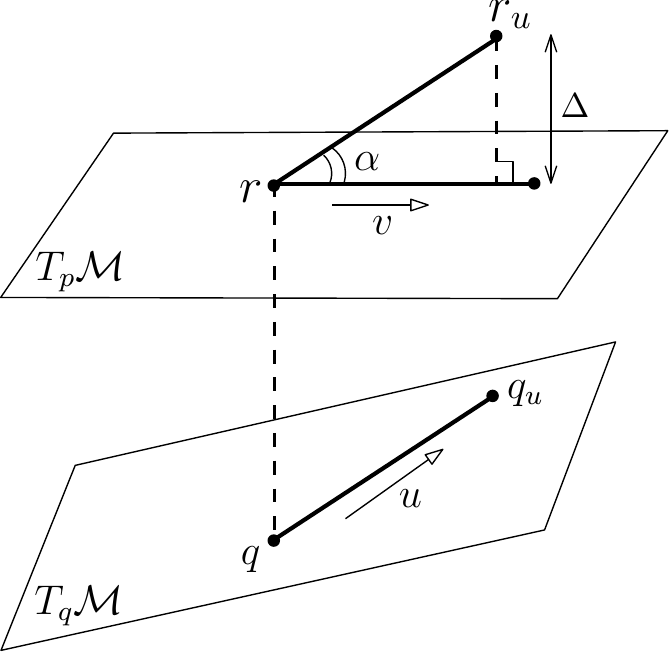}
  \end{center}
%  \vspace{-15pt}
  \caption{In the figure $\alpha = \angle ([r,\, r_{u}], T_{p}\man)$ 
  and $\Delta = \dist{r_{u}}{T_{p}\man}$.}
  \label{Fig-2}
%  \vspace{-25pt}
\end{figure}

Let $r \in T_{p}\man$ be the point closest to $q$ in $T_{p}\man$,
i.e., $\|q-r\| = \dist{q}{T_{p}\man}$, and
let $v$ be an unit vector in $T_{p}\M$ that makes the smallest angle with $u$.
Let $r_{u} = r + t \lfs(q) \cdot u$. Now observe that since 
$\norm{q - r} = \norm{q_{u} - r_{u}}$, we have
\begin{eqnarray}
  \dist{r_{u}}{T_{p}\man} \leq \norm{r_{u}-q_{u}} + \dist{q_{u}}{T_{p}\man}
  = \dist{q}{T_{p}\man} + \dist{q_{u}}{T_{p}\man},
\end{eqnarray}
and the projection of the line segment $[r,\, r_{u}]$ onto $T_{p}\man$ is
parallel to $v$, which implies 
$\angle(u,\, v) = \angle([r,\, r_{u}], T_{p}\man)$.

Using the fact that $\dist{r}{r_{u}} = t \lfs(q) \geq t (1-t)\lfs(p)$
(from Eq.~\eqref{eqn-lfs-p-q-t}),
$\dist{q}{T_{p}\man} \leq \frac{t^{2}}{2}\lfs(p)$ (from
Lemma~\ref{lem:sampling}~(1)), and Eq.~\eqref{eqn-dist-tq-pu}, we get
\begin{eqnarray*}
	\sin \angle (u, v) &=& \sin \angle ([r,\, r_{u}], \, T_{p}\man) \\
	&=& \frac{\dist{r_{u}}{T_{p}\man}}{\dist{r}{r_{u}}} \\
	&\leq& \frac{\dist{q}{T_{p}\man} + \dist{q_{u}}{T_{p}\man}}{\dist{r}{r_{u}}} \\
	&\leq& \frac{t^{2}}{2 \dist{r}{r_{u}}}\left( (2+3t+2t^{2})^{2} + 4(1+t) + 1\right) \lfs(p) \\
	&\leq& t\left( \frac{(2+3t+2t^{2})^{2} + 4t + 5}{2-2t}\right) \\
	&\stackrel{\rm def}{=}& t \, f(t)
%	&=& 4t\, \left( \frac{(1+t)^{3}}{(1-t)^{2}} + \frac{1+t}{1-t} \right) \; < \; 12\,t
\end{eqnarray*}
This completes the proof of Theorem~\ref{theorem-tangent-variation-submanifolds}~(i).
%The above inequality follows from the facts that
%$\eta \leq \frac{2t^{2}}{1-t} \left( \frac{(1+t)^{2}}{(1-t)} + 1\right) \lfs(q)$,
%$\|p-p_{u}\| = t\lfs(p)$, $\lfs(p) \geq \frac{\lfs(q)}{1+t}$ (from
%Eq.~\eqref{eqn-lfs-p-q-t}), $\Theta_{[p,\, p_{u}]} = 1$ and $t \leq 1/12$.

\section{Proof of Theorem~\ref{theorem-tangent-variation-submanifolds}~(ii)}

In this section we will give the proof of 
Theorem~\ref{theorem-tangent-variation-submanifolds}~(ii)
which improves on the 
the asymptotic constant 
in Theorem~\ref{theorem-tangent-variation-submanifolds}~(i)
from $\frac{9}{2}$ to $3$.
We will use Theorem~\ref{theorem-tangent-variation-submanifolds}~(i)
itself to improve the asymptotic constant. 

First observe that we can get an improvement on the 
bound given in Lemma~\ref{lem:sampling}~(2):
\begin{lemma}\label{lem:improved-sampling}
 Let $p \in \man$ and $x \in T_{p}\man$ with 
 $\| p - x \| = t \lfs(p)$ with $t \leq \frac{19}{200}$, 
 then 
 $$
  \dist{x}{\man} \; \leq \; (1-\sqrt{1-t^{2}})\lfs(p) 
  \; = \; \frac{t^{2}\lfs(p)}{1+\sqrt{1-t^{2}}}.
 $$
\end{lemma}
% Using the above result we can get an asymptotic 
% improvement over Theorem~\ref{theorem-tangent-variation-submanifolds}~(i).

% We get above bound in Theorem~\ref{theorem-tangent-variation-submanifolds}~(ii)
By replacing Lemma~\ref{lem:sampling}~(2) with 
Lemma~\ref{lem:improved-sampling} in the proof of 
Theorem~\ref{theorem-tangent-variation-submanifolds}~(i) given in 
Section~\ref{sec:the-proof}, we obtain the improved asymptotic
constant stated in Theorem~\ref{theorem-tangent-variation-submanifolds}~(ii).

%We will now give the details of the proof of Lemma~\ref{lem:improved-sampling}.

% Let $\pi_{p}: \overline{B}(p,r) \cap \man \rightarrow T_{p}\man$ denotes the 
% orthogonal projection map onto $T_{p}\man$ with $r = \frac{\lfs(p)}{10}$. 

We will prove the following stronger result, which implies
Lemma~\ref{lem:improved-sampling}.
\begin{lemma}\label{lem:improved-sampling-2}
 Let $\pi_{p} : \man \rightarrow T_{p}\man$ denote the orthogonal
 projection of $\man$ onto $T_{p}\man$. Then
 \begin{description}
  \item[(i)] 
    $\pi_{p}$ restricted to $\overline{B}_{\mid \man}(p, r)$
    is an embedding, where $r \stackrel{\rm def}{=} {\lfs(p)}/{10}$,
  
  \item[(ii)]
    $\overline{B}_{\mid T_{p}\man}(p, r') \subseteq \pi_{p}(\overline{B}_{\mid \man}(p, r))$
    where $r' \stackrel{\rm def}{=} 19\lfs(p)/200$, and 
  
  \item[(iii)]
    let $x \in T_{p}\man$ with $\| p - x \| = t \lfs(p)$ with $t \leq \frac{19}{200}$, 
    then 
    $\|x - \pi_{p}^{-1}(x) \| \leq \frac{t^{2}\, \lfs(p)}{1+\sqrt{1-t^{2}}}$.
 \end{description}
\end{lemma}

\begin{proof}%[of Lemma~\ref{lem:improved-sampling}] 
 Using the Morse-theory 
 argument from~\cite[Prop.~12]{boissonnat2001}, 
 we can show that 
 $B_{\mid \man}(p,r)$ is an $m$-dimensional topological ball
 where $m$ is the dimension of the manifold $\man$.

 \paragraph{(i)} Observe that for all $x$ and
 $y \in \overline{B}_{\mid \man}(p,r)$, we have from
 Theorem~\ref{theorem-tangent-variation-submanifolds} and
 Lemma~\ref{lem:sampling}~(1)
 \begin{align}\label{eq:x-y-t-p-man}
  \sin \angle ([x, \, y],\, T_{p}\man)
  &\leq \sin \angle ([x, \, y],\, T_{x}\man) + \sin \angle (T_{x}\man,\, T_{p}\man)&\nonumber\\
  &\leq \frac{r}{\lfs(x)} + \frac{6r}{\lfs(p)}&\nonumber\\
  &\leq \frac{r}{\lfs(p)-r} + \frac{6r}{\lfs(p)}&
  \mbox{since $\lfs(x) \geq \lfs(p) - r$}\nonumber \\
  &< 1.&
 \end{align}
 The last inequality follows from the fact that 
 $r = \frac{\lfs(p)}{10}$.
 This implies that $\pi_{p}$ is a homeomorphism 
 between $\overline{B}_{\mid \man}(p,r)$ and $\pi_{p}\left(\overline{B}_{\mid \man}(p,r)\right)$. 
 Since if there exists $x$ and $y \, (\neq x)$ in $\overline{B}(p,r)\cap \man$
 such that $\pi_{p}(x) = \pi_{p}(y)$ then 
 $\angle ([x, \, y], T_{p}\man) = \pi/2$ and that contradicts 
 Eq.~\eqref{eq:x-y-t-p-man}.

 % \begin{figure}%[h]
%   \begin{center}
% %    \quad \quad \quad \quad
%     \includegraphics[width=0.450\textwidth]{Fig-3}
%   \end{center}
%   \caption{Proof of Lemma~\ref{lem:improved-sampling}.}
%   \label{Fig-3}
% \end{figure}

 \paragraph{(ii)}
 We now show that 
 \begin{equation}
   \label{eq:contained.ball}
   \overline{B}_{\mid T_{p}\man}(p, r') \;
   \subseteq \; \pi_{p} \left(\overline{B}_{\mid \man}(p,r)\right)
   \quad \mbox{where $r' = \frac{19\, \lfs(p)}{200}$}.
  \end{equation}
  Since $\overline{B}_{| \man}(p,r)$ is a closed topolgical ball, and
  $\pi_p$ restricted to $\overline{B}_{| \man}(p,r)$ is an embedding,
  it follows that $\pi_p(\overline{B}_{| \man}(p,r))$ is also a
  topolgical ball containing $p$ in its interior, and any point
  $y \in \partial\pi_p(\overline{B}_{| \man}(p,r))$ is the image of a point
  $\pi_p^{-1}(y) \in \partial \overline{B}_{| \man}(p,r)$. Using
  Lemma~\ref{lem:sampling}~(1) we find
 $$
  \| p - y\| \geq  \|p-\pi^{-1}_{p}(y)\| - \|\pi^{-1}_{p}(y) - y\|
  \geq r - \frac{r^{2}}{2 \lfs(p)} = r',
 $$
 and Equation~\eqref{eq:contained.ball} follows.

 % To reach a contradiction let say there exists 
 % $x \in \overline{B}_{\mid T_{p}\man}(p, r') \setminus 
 % \pi_{p} \left(\overline{B}_{\mid \man}(p,r)\right)$,
 % see Fig.~\ref{Fig-3}. 
 % Let $y = {\rm argmax}_{q\in S} \|p-q\|$, where 
 % $S = \pi_{p} \left(\overline{B}_{\mid \man}(p,r)\right) \cap [p,\, x]$. Note that $y \in S$ 
 % and $\|p-y\| < r'$ since 
 % $S$ is compact. Note that $\pi_{p}^{-1}(x) \not\in \partial B_{\mid \man}(p,r)$ 
 % since $\pi_{p}^{-1}(x) \in \partial B_{\mid \man}(p,r)$ would imply,
 % from Lemma~\ref{lem:sampling}~(1),
 % $$
 %  \| p - y\| \geq  \|p-\pi^{-1}_{p}(y)\| - \|\pi^{-1}_{p}(y) - y\|
 %  \geq r - \frac{r^{2}}{2 \lfs(p)} = r'. 
 % $$
 % And if $\pi_{p}^{-1}(x) \in  B_{\mid \man}(p,r)$
 % then we would reach a contradiction since, using the facts that $\pi_{p}$ 
 % restricted to $B_{\mid \man}(p,r)$ is injective 
 % (actually $\pi_{p}$ is injective when restricted to $\overline{B}_{\mid \man}(p,r)$)
 % and $B_{\mid \man}(p,r)$
 % is a $m$-dimensional topological ball, we can show there exists $\epsilon > 0$
 % such that $B_{\mid T_{p}\man}(p, \epsilon) \subset \pi_{p}\left(B_{\mid \man}(p,r)\right)$
 % and that would imply $y$ is not the farthest point from $p$ in $S$.

 \paragraph{(iii)}
 Let $x$ be a point in $T_{p}\man$ with $\|p-x\| = t\lfs(p)$ with $t \leq \frac{19}{200}$.
 Since,  from (ii),
 $\overline{B}_{\mid T_{p}\man}(p,r') \subseteq 
 \pi_{p}\left(\overline{B}_{\mid \man} (p,r)\right)$, 
 it follows that
 there exists $x' \in \overline{B}_{\mid \man}(r, p)$ 
 such that $\pi_{p}(x') = x$. Let $t_{1}= \frac{\| p - x' \|}{\lfs(p)}$. 
 From Lemma~\ref{lem:sampling}~(1) and the fact that $\pi_{p}$ is a 
 projection, we get
 \begin{eqnarray}
  \|p-x'\|^{2} \; = \; 
  t_{1}^{2}\lfs(p)^2 \; = \; \|p-x\|^{2} + \|x-x'\|^{2} 
  \; \leq \; t^{2} \lfs(p)^{2} + \frac{t_{1}^{4}}{4} \lfs(p)^{2}.
 \end{eqnarray}
 Solving the above inequality we get 
 $$
 \frac{t_{1}^{2}}{2} \; \leq \; 1-\sqrt{1-t^{2}} \;
 = \; \frac{t^{2}}{1+\sqrt{1-t^{2}}},
 $$
 and this, together with Lemma~\ref{lem:sampling}~(1), 
 implies
 $$
  \dist{x}{\man} \; \leq \; \dist{x}{x'}  
  \; \leq \; \frac{t_{1}^{2}}{2} \lfs(p)
  \; \leq \; \frac{t^{2}\, \lfs(p)}{1+ \sqrt{1-t^{2}}}.
 $$
\end{proof}

%Note that while proving Lemma~\ref{lem:improved-sampling}, 
%we prove the following stronger result:

\section{Discussion}
\label{sec:discussion}

\begin{figure}%[h]
  \begin{center}
    \quad \quad \quad \quad
    \includegraphics[width=0.60\textwidth]{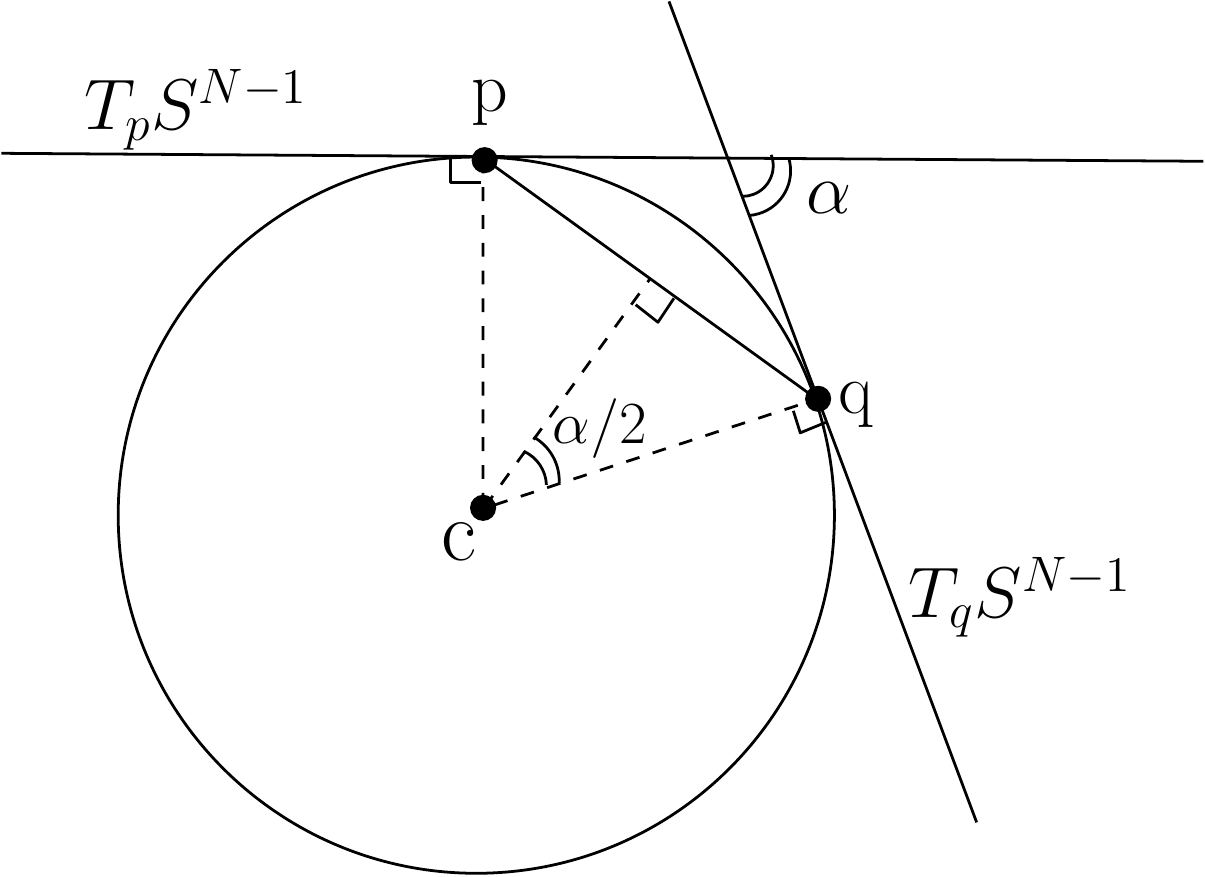}
  \end{center}
  \caption{Tangent variation on a sphere.}
  \label{Fig-1}
\end{figure}

%\begin{wrapfigure}{r}{0.5\textwidth}
%  \vspace{-20pt}
%  \begin{center}
%    \includegraphics[width=0.45\textwidth]{Fig-1}
%  \end{center}
%  \vspace{-10pt}
%  \caption{Tangent variation on a sphere.}
%  \label{Fig-1}
%  \vspace{-20pt}
%\end{wrapfigure}
\paragraph{Lower bound.}
It is easy to see that $O(t)$ bound on the tangent space variation 
on submanifolds is also tight. Consider a $N-1$-dimensional 
unit sphere $\mathbb{S}^{N-1}$ in $\R^{N}$. For all 
$p \in \mathbb{S}^{N-1}$, $\lfs(p) = 1$. Let $p, \, q \in \mathbb{S}^{N-1}$
with $\|p-q\| = t$. Using elementary Euclidean geometry, 
see Fig~\ref{Fig-1},
we can show that 
\begin{eqnarray*}
  \sin\angle (N_{p}\mathbb{S}^{N-1}, N_{q}\mathbb{S}^{N-1})
  = t \sqrt{1 - \frac{t^{2}}{4}}
  = \Omega(t) 
\end{eqnarray*}
Using the above equation, and the fact that 
$\angle (T_{p}\mathbb{S}^{N-1}, T_{q}\mathbb{S}^{N-1})
= \angle (N_{p}\mathbb{S}^{N-1}, N_{q}\mathbb{S}^{N-1})$,
see~\cite[Lem.~2.1]{DBLP:journals/dcg/BoissonnatG14}, we get the lower bound.

%\begin{figure}
%  \begin{center}
%    \includegraphics[width=0.50\textwidth]{Fig-1}
%  \end{center}
%  \caption{Tangent variation on $(N-1)$-dimension unit sphere $\mathbb{S}^{N-1}$.}
%\end{figure}

\paragraph{Regarding the constants.} 
The asymptotic constant we obtained 
in Theorem~\ref{theorem-tangent-variation-submanifolds} is $3$,
%For $t \leq 1/10$, we have $f(t) < 6$, 
unlike the 2-dimensional case where it is $1$, and we expect 
the asymptotic constant should be closer to $1$.

%\paragraph{Acknowledgements.} 

\section*{Acknowledgement}

This research has been partially supported by the 7th Framework
Programme for Research of the European Commission, under FET-Open
grant number 255827 (CGL Computational Geometry Learning). Partial 
support has also been provided by the Advanced Grant of the European 
Research Council GUDHI (Geometric Understanding in Higher Dimensions).

Arijit Ghosh is supported by the Indo-German Max Planck Center for
Computer Science (IMPECS).

\bibliographystyle{alpha}

\bibliography{biblio}

\begin{thebibliography}{NSW08}

\bibitem[AB99]{AmentaBern}
N.~Amenta and M.~Bern.
\newblock Surface reconstruction by {V}oronoi filtering.
\newblock {\em Discrete \& Computational Geometry}, 22:481--504, 1999.

\bibitem[AD14]{DBLP:journals/corr/AmentaD14}
N.~Amenta and T.~K. Dey.
\newblock Normal variation for adaptive feature size.
\newblock {\em CoRR}, abs/1408.0314, 2014.

\bibitem[BC01]{boissonnat2001}
J.-D. Boissonnat and F.~Cazals.
\newblock Natural {N}eighbour {C}oordinates of {P}oints on a {S}urface.
\newblock {\em Computational Geometry: Theory and Applications},
  19(2):155--173, 2001.

\bibitem[BG14]{DBLP:journals/dcg/BoissonnatG14}
J.-D. Boissonnat and A.~Ghosh.
\newblock {Manifold Reconstruction Using Tangential Delaunay Complexes}.
\newblock {\em Discrete {\&} Computational Geometry}, 51(1):221--267, 2014.

\bibitem[BSW09]{DBLP:conf/soda/BelkinSW09}
M.~Belkin, J.~Sun, and Y.~Wang.
\newblock Constructing {L}aplace operator from point clouds in
  \emph{R}\({}^{\mbox{\emph{d}}}\).
\newblock In {\em Proceedings of the Twentieth Annual {ACM-SIAM} Symposium on
  Discrete Algorithms (SODA)}, pages 1031--1040, 2009.

\bibitem[Gho12]{ghosh:hal-01095861}
A.~Ghosh.
\newblock {\em {Piecewise linear reconstruction and meshing of submanifolds of
  Euclidean space}}.
\newblock PhD thesis, {INRIA Sophia Antipolis} \& {Universit{\'e} de Nice
  Sophia Antipolis}, May 2012.

\bibitem[GW04]{DBLP:journals/dcg/GiesenW04}
J.~Giesen and U.~Wagner.
\newblock {Shape Dimension and Intrinsic Metric from Samples of Manifolds}.
\newblock {\em Discrete {\&} Computational Geometry}, 32(2):245--267, 2004.

\bibitem[NSW08]{nsw-fhswhc-04}
P.~Niyogi, S.~Smale, and S.~Weinberger.
\newblock {Finding the {H}omology of {S}ubmanifolds with {H}igh {C}onfidence
  from {R}andom samples}.
\newblock {\em Discrete \& Computational Geometry}, 39(1-3):419--441, 2008.

\end{thebibliography}

\end{document}